\newcounter{myctr}

\documentclass{ws-acs}

\usepackage{latexsym}
\usepackage{epsfig}
\usepackage{amssymb}
\usepackage{tipa}

\begin{document}

\markboth{Mukherjee \bgroup et al.\egroup}{Rediscovering the Co-occurence Principles of Vowel Inventories}

\catchline{}{}{}{}{}
%
%%%%%%%%%%%%%%%%%%%%%%%%%%%%%%%%%%%%%%%%%%%%%%%%%%%%%%%%%%%%%%%%%%%%

\title{Rediscovering the Co-occurrence\\ Principles of Vowel Inventories:\\ A Complex Network Approach}

\author{Animesh Mukherjee, Monojit Choudhury, Anupam Basu, Niloy Ganguly}
\address{Department of Computer Science and Engineering,\\ Indian Institute of Technology, Kharagpur}
\author{Shamik RoyChowdhury}
\address{Department of Information Technology,\\ Heritage Institute of Technology, Kolkata}

\maketitle

\begin{history}
\received{(received date)}
\revised{(revised date)}
%\accepted{(Day Month Year)}
%\comby{(xxxxxxxxxx)}
\end{history}

\begin{abstract}
% Text of abstract
In this work, we attempt to capture patterns of co-occurrence across vowel systems and at the same time figure out the %%@
nature of the force leading to the emergence of such patterns. For this purpose we define a weighted network where the %%@
vowels are the nodes and an edge between two nodes (read vowels) signify their co-occurrence likelihood over the vowel %%@
inventories. Through this network we identify communities of vowels, which essentially reflect their patterns of %%@
co-occurrence across languages. We observe that in the assortative vowel communities the constituent nodes (read %%@
vowels) are largely uncorrelated in terms of their features and show that they are formed based on the principle of %%@
maximal perceptual contrast. However, in the rest of the communities, strong correlations are reflected among the %%@
constituent vowels with respect to their features indicating that it is the principle of feature economy that binds %%@
them together. We validate the above observations by proposing a quantitative measure of perceptual contrast as well %%@
as feature economy and subsequently comparing the results obtained due to these quantifications with those where we %%@
assume that the vowel inventories had evolved just by chance.

\end{abstract}

\keywords{Vowels; complex network; community structure; feature entropy.}  

\section{Introduction}
Linguistic research has documented a wide range of regularities across the sound systems of the world's %%@
languages~\cite{Boer:00,Choudhury:06,Lindblom:72,Lindblom:86,Mukherjee:06,Mukh:06}. Functional phonologists argue that %%@
such regularities are the consequences of certain general principles like {\em maximal perceptual %%@
contrast}\footnote{Maximal perceptual contrast, is desirable between the phonemes of a language for proper perception %%@
of each individual phoneme in a noisy environment}~\cite{Lindblom:72}, {\em ease of articulation}\footnote{Ease of %%@
articulation requires that the sound systems of all languages are formed of certain universal (and highly frequent) %%@
sounds.}~\cite{Boer:00,Lindblom:88}, and {\em ease of learnability}\footnote{Ease of learnability is required so that %%@
a speaker can learn the sounds of a language with minimum effort.}~\cite{Boer:00}. In the study of vowel systems the %%@
optimizing principle, which has a long tradition~\cite{Jakobson:41,Wang:68} in linguistics, is maximal perceptual %%@
contrast. A number of numerical studies based on this principle have been reported in %%@
literature~\cite{Lindblom:72,Lindblom:86,Schwartz:97}. Of late, there have been some attempts to explain the vowel %%@
systems through multi agent simulations~\cite{Boer:00} and genetic algorithms~\cite{Ke:03}; all of these experiments %%@
also use the principle of perceptual contrast for optimization purposes. 

An exception to the above trend is a school of linguists~\cite{Boersma:98,Clements:04} who argue that perceptual %%@
contrast-based theories fail to account for certain fundamental aspects such as the patterns of co-occurrence of %%@
vowels based on similar acoustic/articulatory {\em features}\footnote{In linguistics, features are the elements, which %%@
distinguish one phoneme from another. The features that describe the vowles can be broadly categorized into three %%@
different classes namely the {\em height}, the {\em backness} and the {\em roundedness}. Height refers to the vertical %%@
position of the tongue relative to either the roof of the mouth or the aperture of the jaw. Backness refers to the %%@
horizontal tongue position during the articulation of a vowel relative to the back of the mouth. Roundedness refers to %%@
whether the lips are rounded or not during the articulation of a vowel. There are however still more possible features %%@
of vowel quality, such as the velum position (e.g., nasality), type of vocal fold vibration (i.e., phonation), and %%@
tongue root position (i.e., secondary place of articulation).} observed across the vowel inventories. Instead, they %%@
posit that the observed patterns, especially found in larger size inventories~\cite{Boersma:98}, can be explained only %%@
through the principle of {\em feature economy}~\cite{Groot:31,Martinet:55}. According to this principle, languages %%@
tend to maximize the combinatorial possibilities of a few distinctive features to generate a large number of sounds. 

\begin{figure}
%\begin{center}
\centerline{\psfig{file=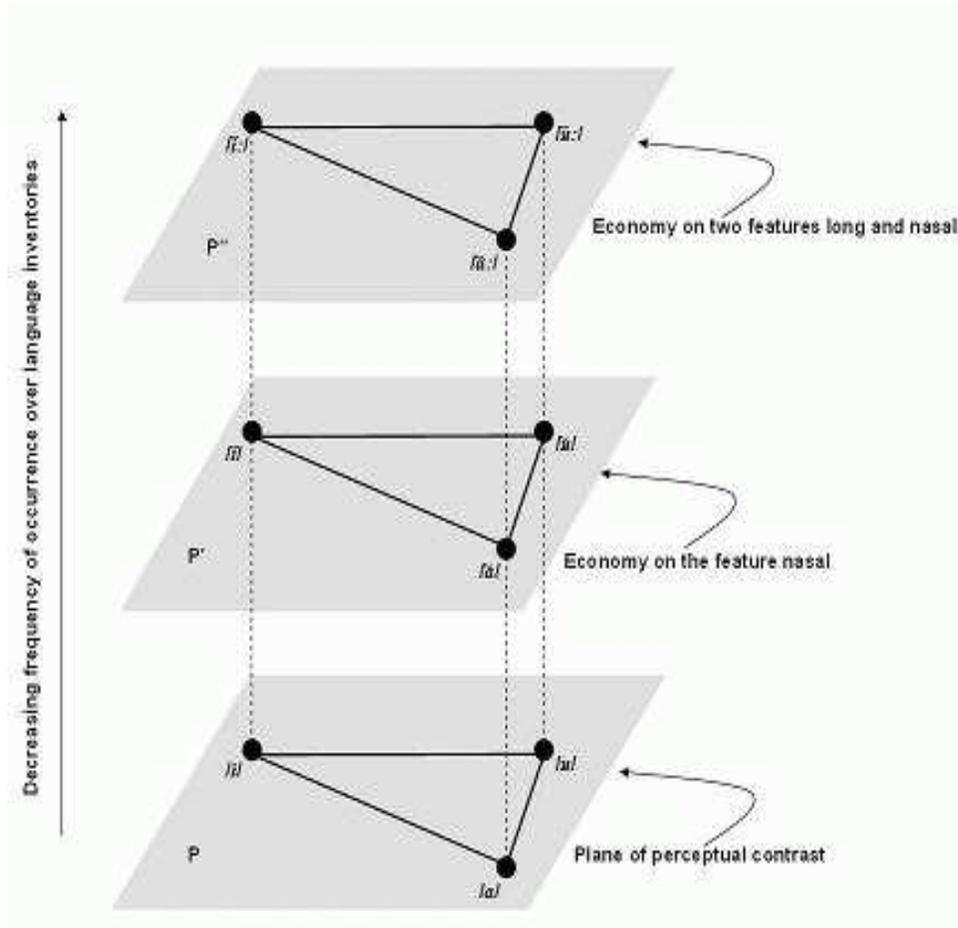,width=5in}}
\caption{The organizational principles of the vowels (in decreasing frequency of occurrence) indicated through %%@
different hypothetical planes.}
\label{vlplane}
%\end{center}
\end{figure}

The aforementioned ideas can be possibly linked together through the example illustrated by Figure~\ref{vlplane}. As %%@
shown in the figure, the initial plane $P$ constitutes of a set of three very frequently occurring vowels /$i$/, /$a$/ %%@
and /$u$/, which usually make up the smaller inventories and do not have any single feature in common. Thus, smaller %%@
inventories are quite likely to have vowels that exhibit a large extent of contrast in their constituent features. %%@
However, in bigger inventories, members from the higher planes ($P^\prime$ and $P^{\prime\prime}$) are also present %%@
and they in turn exhibit feature economy. For instance, in the plane $P^\prime$ comprising of the set of vowels %%@
/$\textipa{\~i}$/, /$\textipa{\~a}/$, /$\textipa{\~u}$/, we find a nasal modification applied equally on all the three %%@
members of the set. This is actually indicative of an economic behavior that the larger inventories show while %%@
choosing a new feature in order to reduce the learnability effort of the speakers. The third plane $P^{\prime\prime}$ %%@
reinforces this idea by showing that the larger the size of the inventories the greater is the urge for this economy %%@
in the choice of new features. Another interesting facet of the figure are the relations that exist across the planes %%@
(indicated by the broken lines). All these relations are representative of a common linguistic concept of {\em %%@
robustness}~\cite{Clements:04} in which one frequently occurring vowel (say /$i$/) implies the presence of the other %%@
(and not vice versa) less frequently occurring vowel (say /$\textipa{\~i}$/) in a language inventory. These %%@
cross-planar relations are also indicative of feature economy since all the features present in the frequent vowel %%@
(e.g., /$i$/) are also shared by the less frequent one (e.g., /$\textipa{\~i}$/). In summary, while the basis of %%@
organization of the vowel inventories is perceptual contrast as indicated by the plane $P$ in Figure~\ref{vlplane}, %%@
economic modifications of the perceptually distinct vowels takes place with the increase in the inventory size (as %%@
indicated by the planes $P^\prime$ and $P^{\prime\prime}$ in Figure~\ref{vlplane}). 

In this work we attempt to corroborate the above conjecture by automatically capturing the patterns of co-occurrence %%@
that are prevalent {\em in} and {\em across} the planes illustrated in Figure~\ref{vlplane}. We also present a %%@
quantitative measure of the driving forces that lead to the emergence of such patterns and show that the real %%@
inventories are significantly better in terms of this measure than expected. In order to do so, we define the ``{\bf %%@
Vo}wel-Vowel {\bf Net}work" or {\bf VoNet}, which is a weighted network where the vowels are the nodes and an edge %%@
between two nodes (read vowels) signify their co-occurrence likelihood over the vowel inventories. We conduct %%@
community structure analysis of different versions of VoNet in order to capture the patterns of co-occurrence in and %%@
across the planes $P$, $P^\prime$ and $P^{\prime\prime}$ shown in Figure~\ref{vlplane}. The plane $P$ consists of the %%@
communities, which are formed of those vowels that have a very high frequency of occurrence (usually {\em %%@
assortative}~\cite{Newman:03} in nature). We observe that the constituent nodes (read vowels) of these assortative %%@
vowel communities are largely uncorrelated in terms of their features and quantitatively show that they indeed exhibit %%@
a higher than expected level of perceptual contrast. On the other hand, the communities obtained from VoNet, in which %%@
the links between the assortative nodes are absent, corresponds to the co-occurrence patterns of the planes $P^\prime$ %%@
and $P^{\prime\prime}$. In these communities, strong correlations are reflected among the constituent vowels with %%@
respect to their features and they indeed display a significantly better feature economy than it could have been by %%@
random chance. Moreover, the co-occurrences across the planes can be captured by the community analysis of VoNet where %%@
only the connections between the assortative and the non-assortative nodes, with the non-assortative node co-occurring %%@
very frequently with the assortative one, are retained while the rest of the connections are filtered out. We also %%@
show that these communities again exhibit a significantly higher feature economy than feasible by chance.      

This article is organized as follows: Section~\ref{setup} describes the experimental setup in order to explore the %%@
co-occurrence principles of the vowel inventories. In this section we formally define VoNet, outline its construction %%@
procedure, present a community-finding algorithm, and also present a quantitative definition for maximal perceptual %%@
contrast as well as feature economy. In section~\ref{results} we report the experiments performed to obtain the %%@
community structures, which are representative of the co-occurrence patterns in and across the planes discussed above. %%@
We also report results where we measure the driving forces that lead to the emergence of such patterns and show that %%@
the real inventories are substantially better in terms of this measure than those where the inventories are assumed to %%@
have evolved by chance. Finally, we conclude in section~\ref{conc} by summarizing our contributions, pointing out some %%@
of the implications of the current work and indicating the possible future directions.   

\section{Experimental Setup}\label{setup}

In this section we systematically develop the experimental setup in order to investigate the co-occurrence principles %%@
of the vowel inventories. For this purpose, we formally define VoNet, outline its construction procedure, describe a %%@
community-finding algorithm to decompose VoNet to obtain the community structures, and define the metrics required in %%@
order to explore the co-occurrence principles of the observed communities.   

\subsection{Definition and Construction of VoNet}

\begin{figure}
\begin{center}
\framebox[3.5in]{
\psfig{file=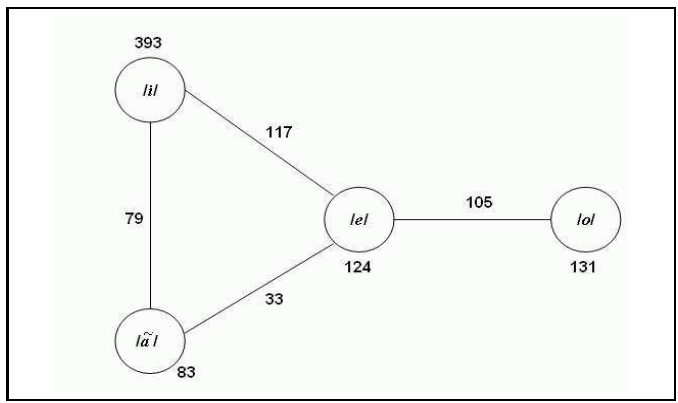,width=3in}
}
\caption{A partial illustration of the nodes and edges in VoNet. The labels of the nodes denote the vowels represented %%@
in IPA (International Phonetic Alphabet). The numerical values against the edges and nodes represent their %%@
corresponding weights. For example /$i$/ occurs in 393 languages; /$e$/ occurs in 124 languages while they co-occur in %%@
117 languages.}
\label{graph}
\end{center}
\end{figure}

{\bf Definition of VoNet:} We define VoNet as a network of vowels, represented as G = $\langle$ V$_V$, E $\rangle$ %%@
where V$_V$ is the set of nodes labeled by the vowels and E is the set of edges occurring in VoNet. There is an edge %%@
$e$ $\in$ E between two nodes, if and only if there exists one or more language(s) where the nodes (read vowels) %%@
co-occur. The weight of the edge $e$ (also {\em edge-weight}) is the number of languages in which the vowels connected %%@
by $e$ co-occur. The weight of a node $u$ (also {\em node-weight}) is the number of languages in which the vowel %%@
represented by $u$ occurs. In other words, if a vowel $v_i$ represented by the node $u$ occurs in the inventory of $n$ %%@
languages then the node-weight of $u$ is assigned the value $n$. Also if the vowel $v_j$ is represented by the node %%@
$v$ and there are $w$ languages in which vowels $v_i$ and $v_j$ occur together then the weight of the edge connecting %%@
$u$ and $v$ is assigned the value $v$. Figure~\ref{graph} illustrates this structure by reproducing some of the nodes %%@
and edges of VoNet.\\

\begin{figure*}
%\begin{center}
\centerline{\psfig{file=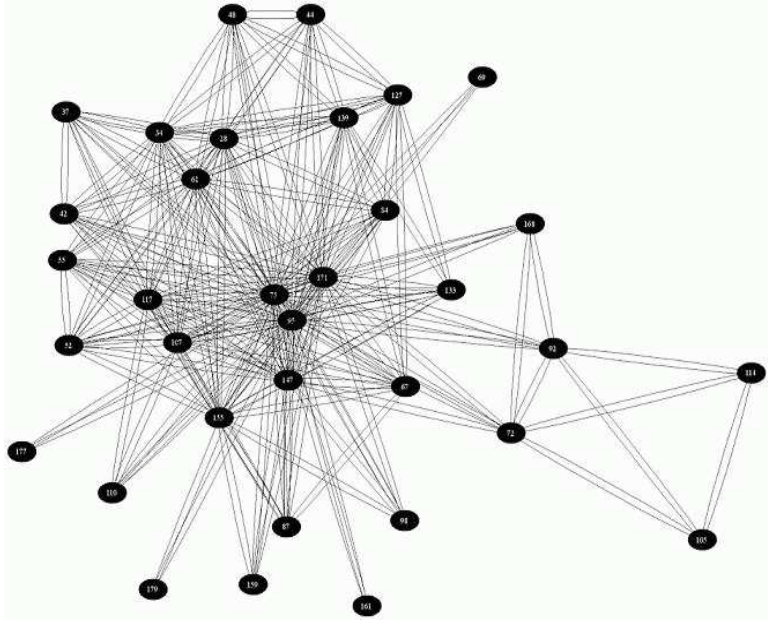,width=4in}}
\caption{A partial illustration of VoNet. All edges in this figure have an edge-weight greater than or equal to 15. %%@
The number on each node corresponds to a particular vowel. For instance, node number 72 corresponds to %%@
\textipa{/\~i/}.}
\label{ph10}
%\end{center}
\end{figure*}

{\bf Construction of VoNet:} Many typological %%@
studies~\cite{Choudhury:06,Hinskens:03,Ladefoged:96,Lindblom:88,Mukherjee:06,Mukh:06} of segmental inventories have %%@
been carried out in past on the UCLA Phonological Segment Inventory Database (UPSID)~\cite{Maddieson:84}. Currently %%@
UPSID records the sound inventories of 451 languages covering all the major language families of the world. In this %%@
work we have therefore used UPSID comprising of these 451 languages and 180 vowels found across them, for constructing %%@
VoNet. Consequently, the set V$_V$ comprises of 180 elements (nodes) and the set E comprises of 3135 elements (edges). %%@
Figure~\ref{ph10} presents a partial illustration of VoNet as constructed from UPSID.

\subsection{Finding Community Structures}

%radicchi algo
We attempt to identify the communities appearing in VoNet by the extended Radicchi \bgroup et al. %%@
\egroup~\cite{Rad:03} algorithm for weighted networks as introduced by us in an earlier article~\cite{Mukherjee:06}. %%@
The basic idea is that if the weights on the edges forming a triangle (loops of length three) are comparable then the %%@
group of vowels represented by this triangle highly occur together rendering a pattern of co-occurrence while if these %%@
weights are not comparable then there is no such pattern. In order to capture this property we define a strength %%@
metric $S$ for each of the edges of VoNet as follows. Let the weight of the edge ($u$,$v$), where $u$, $v$ $\in$ %%@
V$_C$, be denoted by $w_{uv}$. We define $S$ as,         
\begin{equation}
S =  \frac{w_{uv}}{\sqrt{\sum_{i \in V_C - \{u,v\}}{(w_{ui} - w_{vi})}^2}}    
\end{equation} 
if $\sqrt{\sum_{i \in V_C - \{u,v\}}{(w_{ui} - w_{vi})}^2} > 0$ else $S = \infty$. The denominator in this expression %%@
essentially tries to capture whether or not the weights on the edges forming triangles are comparable (the higher the %%@
value of $S$ the more comparable the weights are). The network can be then decomposed into clusters or communities by %%@
removing edges that have $S$ less than a specified threshold (say $\eta$).

At this point it is worthwhile to clarify the significance of a vowel community. A community of vowels actually refers %%@
to a set of vowels which occur together in the language inventories very frequently. In other words, there is a higher %%@
than expected probability of finding a vowel $v$ in an inventory which already hosts the other members of the %%@
community to which $v$ belongs. For instance, if /$i$/, /$a$/ and /$u$/ form a vowel community and if /$i$/ and /$a$/ %%@
are present in any inventory then there is a very high chance that the third member /$u$/ is also present in the %%@
inventory. 

\subsection{Definition of the Metrics}

Once the communities are obtained through the algorithm discussed earlier the next important task is to analyze them %%@
so as to capture the binding force that keeps them together. For this purpose, we need to have a quantitative measure %%@
for perceptual contrast as well as feature economy. In order to establish that the above forces really play a role in %%@
the emergence of the communities, we also need to compare and show that the communities are much better in terms of %%@
this measure than it would have been if the vowel inventories had evolved by chance. In the rest of this section we %%@
detail out the metric for quantification as well as the metric for comparison. 
 
\subsubsection{Metric for Quantification}

For a community $C$ of size $N$ let there be $p_f$ vowels, which have a particular feature $f$ (where $f$ is assumed %%@
to be boolean in nature)\footnote{There are 28 such boolean features that are found across the vowel systems recorded %%@
in UPSID.} in common and $q_f$ other vowels, which lack the feature $f$. Thus, the probability that a particular vowel %%@
chosen uniformly at random from C has the feature $f$ is $\frac {p_f}{N}$ and the probability that the vowel lacks the %%@
feature $f$ is $\frac {q_f}{N}$ (=1--$\frac{p_f}{N}$). If $F$ be the set of all features present in the vowels in $C$ %%@
then {\em feature entropy} $F_E$ can be defined as
\begin{equation}
F_E = \sum_{f \in F}(- \frac {p_f}{N}\log{\frac{p_f}{N}} - \frac{q_f}{N}\log{\frac {q_f}{N}})        
\end{equation} 
$F_E$ is essentially the measure of the number of bits that are required to communicate the information about the %%@
entire community $C$ through a channel.\\

\begin{figure}
\begin{center}
\framebox{
\psfig{file=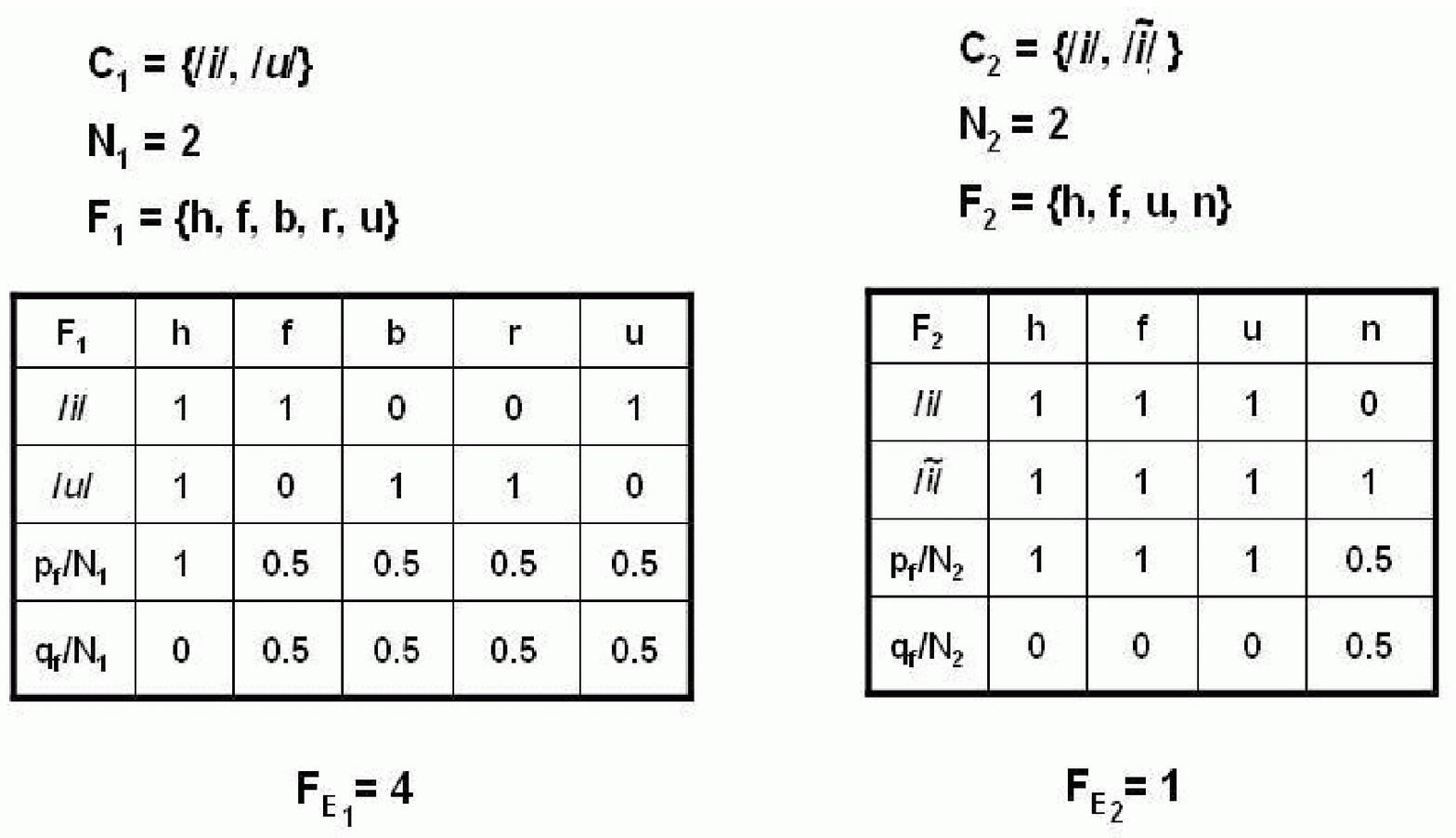,width=5in}
}
\caption{$F_E$ for the two different communities $C_1$ and $C_2$. The letters {\bf h}, {\bf f}, {\bf b}, {\bf r}, {\bf %%@
u}, {\bf l}, and {\bf n} stand for the features high, front, back, rounded, unrounded, and nasalized respectively.}
\label{mc_ex}
\end{center}
\end{figure}

{\bf Capturing Perceptual Contrast:} If $C$ comprises of a set of perceptually distinct vowels, then larger number of %%@
bits should be required to communicate the information about $C$ over the transmission channel since in this case the %%@
set of features that constitute the vowels are more in number. Therefore, the higher the perceptual contrast the %%@
higher is the feature entropy. The idea is illustrated through the example in Figure~\ref{mc_ex}. In the figure, $F_E$ %%@
exhibited by the community $C_1$ is higher than that of the community $C_2$, since the set of vowels in $C_1$ are %%@
perceptually more distinct than those in $C_2$.\\

\begin{figure}
\begin{center}
\framebox{
\psfig{file=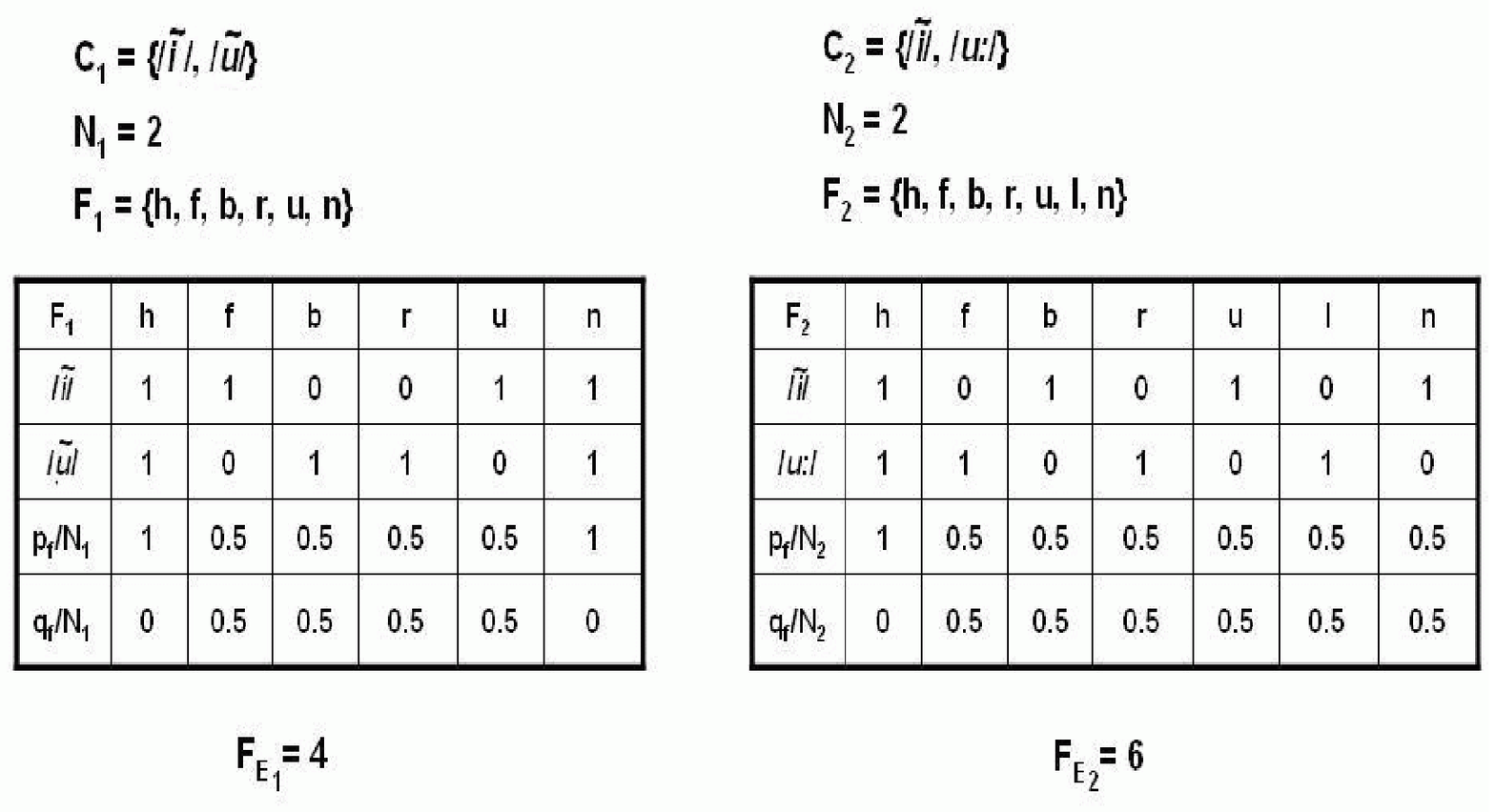,width=5in}
}
\caption{$F_E$ for the two different communities $C_1$ and $C_2$. The letters {\bf h}, {\bf f}, {\bf b}, {\bf r}, {\bf %%@
u}, {\bf l}, and {\bf n} stand for the features high, front, back, rounded, unrounded, long, and nasalized %%@
respectively.}
\label{fe_ex}
\end{center}
\end{figure}

{\bf Capturing Feature Economy:} To have more information conveyed using a fewer number of bits, maximization of the %%@
combinatorial possibilities of the features used by the constituent vowels in the community $C$ is needed, which is %%@
precisely the prediction made by the principle of feature economy. Therefore the lower the feature entropy the higher %%@
is the feature economy. In fact, it is due to this reason that in Figure~\ref{fe_ex}, $F_E$ exhibited by the community %%@
$C_1$ is lower than that of the community $C_2$, since in $C_1$ the combinatorial possibilities of the features is %%@
better utilized by the vowels than in $C_2$. 

\subsubsection{Metric for Comparison}

For the purpose of the comparison as discussed earlier, we construct a random version of VoNet, namely VoNet$_{rand}$. %%@
Let the frequency of occurrence for each vowel $v$ in UPSID  be denoted by $f_v$. Let there be 451 bins each %%@
corresponding to a language in UPSID. $f_v$ bins are then chosen uniformly at random and the vowel $v$ is packed into %%@
these bins. Thus the vowel inventories of the 451 languages corresponding to the bins are generated. In such randomly %%@
constructed inventories the effect of none of the forces (perceptual contrast or feature economy) should be prevalent %%@
as there is no strict co-occurrence principle that plays a role in the inventory construction. Therefore these %%@
inventories should show a feature entropy no better than expected by random chance and hence can act as a baseline for %%@
all our experiments reported in the following section. VoNet$_{rand}$ can be then constructed from these new vowel %%@
inventories similarly as VoNet. The method for the construction is summarized in Algorithm~1.

\begin{algorithm}{Algorithm to construct VoNet$_{rand}$}\\
 {\bf for} {\em each vowel v}\\
  \{
\\
   \hspace*{20pt}{\bf for} {\em i = 1 to $f_v$}\\
   \hspace*{20pt}\{\\
    \hspace*{40pt}Choose one of the 451 bins, corresponding to the languages in UPSID, \hspace*{40pt}uniformly at %%@
random;\\
	\hspace*{40pt}Pack the vowel $v$ into the bin so chosen if it has not been already \hspace*{40pt}packed into this %%@
bin earlier;\\  
   \hspace*{20pt}\}\\ 
  \} 
  \\
Construct VoNet$_{rand}$, similarly as VoNet, from the new vowel inventories (each bin corresponds to a new %%@
inventory);
\end{algorithm}
\label{rand}

\section{Experiments and Results}\label{results}

In this section we describe the experiments performed and the results obtained from the analysis of VoNet. In order to %%@
find the co-occurrence patterns in and across the planes of Figure~\ref{vlplane} we define three versions of VoNet %%@
namely VoNet$_{assort}$, VoNet$_{rest}$ and VoNet$_{rest^\prime}$. The construction procedure for each of these %%@
versions are presented below.\\

\begin{figure}
%\begin{center}
\centerline{\psfig{file=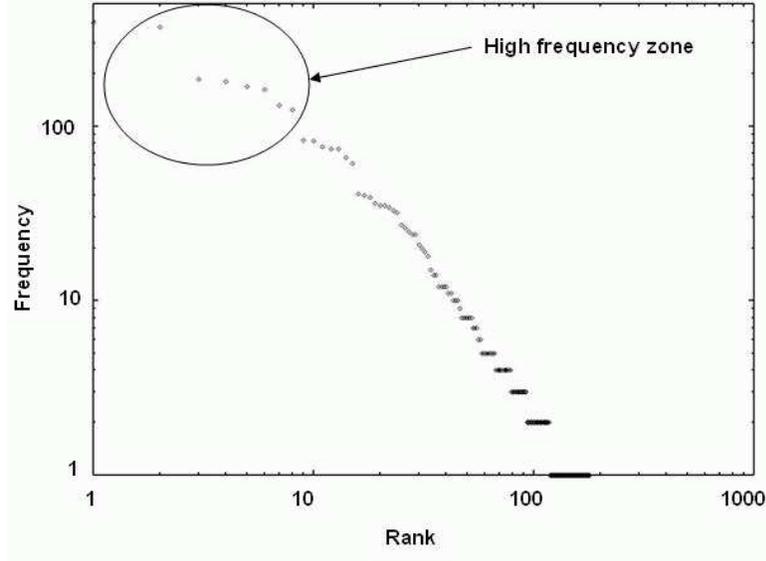,width=4in}}
\caption{The frequency (y-axis) versus rank (x-axis) curve in log-log scale illustrating the distribution of the %%@
occurrence of the vowels over the language inventories of UPSID.}
\label{fq_dis}
%\end{center}
\end{figure}

{\bf Construction of VoNet$_{assort}$:} VoNet$_{assort}$ comprises of the assortative\footnote{The term ``assortative %%@
node" here refers to the nodes having a very high node-weight.} nodes having node-weights above 120 (i.e, vowels %%@
occurring in more than 120 languages in UPSID), along with only the edges inter-connecting these nodes. The rest of %%@
the nodes (having node-weight less than 120) and edges are removed from the network. We make a choice of this %%@
node-weight for classifying the assortative nodes from the non-assortative ones by observing the distribution of the %%@
occurrence frequency of the vowels illustrated in Figure~\ref{fq_dis}. The curve shows the frequency of a vowel %%@
(y-axis) versus the rank of the vowel according to this frequency (x-axis) in log-log scale. The high frequency zone %%@
(marked by a circle in the figure) can be easily distinguished from the low-frequency one since there is distinct gap %%@
featuring between the two in the curve.\\  

\begin{figure}
%\begin{center}
\centerline{\psfig{file=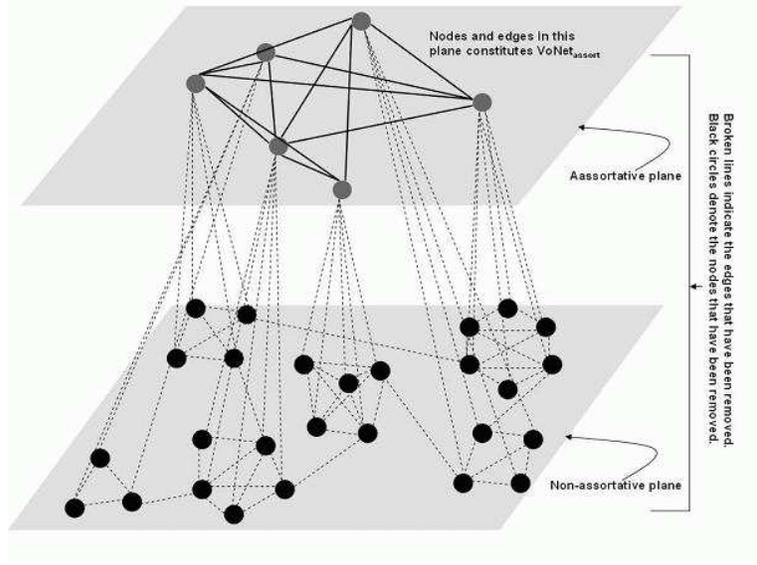,width=4in}}
\caption{The construction procedure of VoNet$_{assort}$ from VoNet.}
\label{VoNetass}
%\end{center}
\end{figure}

Figure~\ref{VoNetass} illustrates how VoNet$_{assort}$ is constructed from VoNet. Presently, the number of nodes in %%@
VoNet$_{assort}$ is 9 and the number of edges is 36. 

\begin{figure}
%\begin{center}
\centerline{\psfig{file=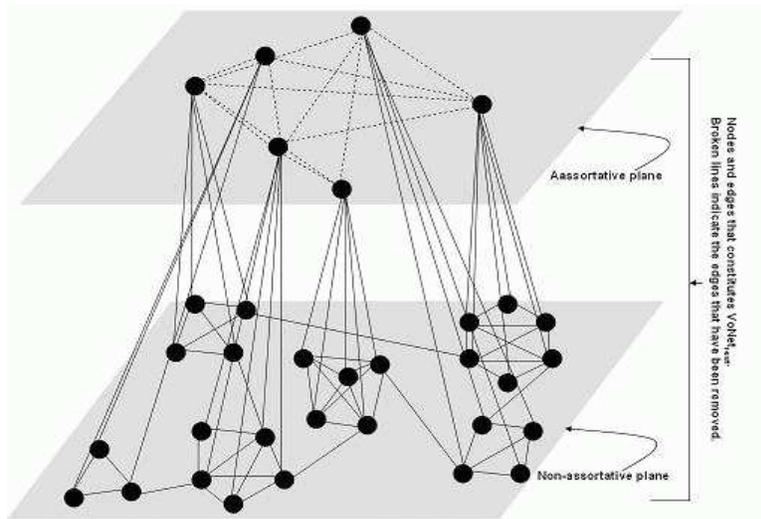,width=4in}}
\caption{The construction procedure of VoNet$_{rest}$ from VoNet.}
\label{VoNetres}
%\end{center}
\end{figure}

{\bf Construction of VoNet$_{rest}$:} VoNet$_{rest}$ comprises of all the nodes as that of VoNet. It also has all the %%@
edges of VoNet except for those edges that inter-connect the assortative nodes. Figure~\ref{VoNetres} shows how %%@
VoNet$_{rest}$ can be constructed from VoNet. The number of nodes and edges in VoNet$_{rest}$ are 180 and %%@
1293\footnote{We have neglected nodes with node-weight less than 3 since these nodes correspond to vowels that occur %%@
in less than 3 languages in UPSID and the communities they form are therefore statistically insignificant.} %%@
respectively.\\  

\begin{figure}
%\begin{center}
\centerline{\psfig{file=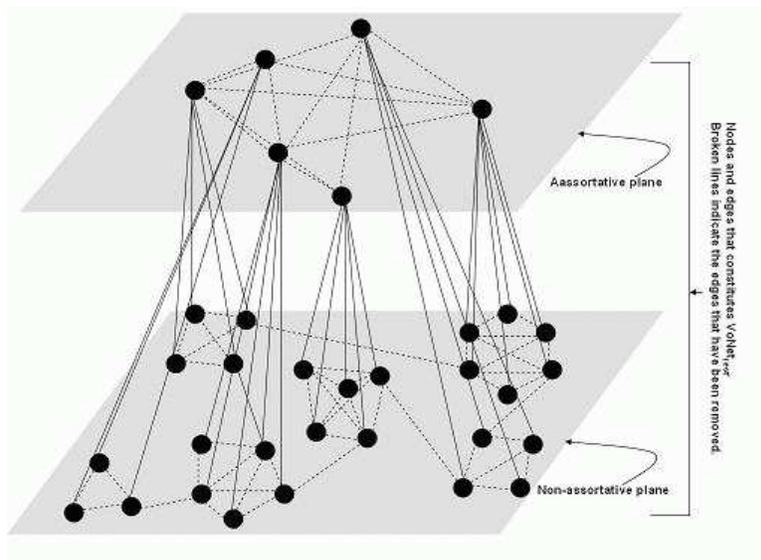,width=4in}}
\caption{The construction procedure of VoNet$_{rest^\prime}$ from VoNet.}
\label{VoNetresd}
%\end{center}
\end{figure}

{\bf Construction of VoNet$_{rest^\prime}$:} VoNet$_{rest^\prime}$ again comprises of all the nodes as that of VoNet. %%@
It consists of only the edges that connect an assortative node with a non-assortative one if the non-assortative node %%@
co-occurs more than ninety five percent of times with the assortative nodes. The basic idea behind such a construction %%@
is to capture the co-occurrence patterns based on robustness~\cite{Clements:04} (discussed earlier in the introductory %%@
section) that actually defines the cross-planar relationships in Figure~\ref{vlplane}. Figure~\ref{VoNetresd} shows %%@
how VoNet$_{rest^\prime}$ can be constructed from VoNet. The number of nodes in VoNet$_{rest^\prime}$ is 180 while the %%@
number of edges is 114\footnote{The network does not get disconnected due to this construction since, there is always %%@
a small fraction of edges that run between assortative and low node-weight non-assortative nodes of otherwise disjoint %%@
groups.}.

We separately apply the community-finding algorithm (discussed earlier) on each of VoNet$_{assort}$, VoNet$_{rest}$ %%@
and VoNet$_{rest^\prime}$ in order to obtain the respective vowel communities. We can obtain different sets of %%@
communities by varying the threshold $\eta$. A few assortative vowel communities (obtained from VoNet$_{assort}$) are %%@
noted in Table~\ref{assort}. Some of the communities obtained from VoNet$_{rest}$ are presented in Table~\ref{rest}. %%@
We also note some of the communities obtained from VoNet$_{rest^\prime}$ in Table~\ref{restd}.  

% Table created by WinTeX 2000: 2 Columns x 3 Rows.

\begin{table}\centering
\caption{Assortative vowel communities. The contrastive features separated by slashes (/) are shown within %%@
parentheses. Comma-separated entries represent the features that are in use from the three respective classes namely %%@
the height, the backness, and the roundedness.}
\begin{tabular}{|l|l|}
%Row: 1
\cline{1-2}
\vbox to1.88ex{\vspace{1pt}\vfil\hbox to12.80ex{\hfil Community\hfil}} & 
\vbox to1.88ex{\vspace{1pt}\vfil\hbox to63.20ex{\hfil Features in Contrast\hfil}} \\

%Row: 2
\cline{1-2}
\vbox to1.88ex{\vspace{1pt}\vfil\hbox to12.80ex{\hfil /i/, /a/, /u/\hfil}} & 
\vbox to1.88ex{\vspace{1pt}\vfil\hbox to63.20ex{\hfil (low/high), (front/central/back), (unrounded/rounded)\hfil}} \\

%Row: 3
\cline{1-2}
\vbox to1.88ex{\vspace{1pt}\vfil\hbox to12.80ex{\hfil /e/, /o/\hfil}} & 
\vbox to1.88ex{\vspace{1pt}\vfil\hbox to63.20ex{\hfil (higher-mid/mid), (front/back), (unrounded/rounded)\hfil}} \\

\cline{1-2}
\end{tabular}
\label{assort}
\end{table}

% Table created by WinTeX 2000: 2 Columns x 4 Rows.
\begin{table}\centering
\caption{Some of the vowel communities obtained from VoNet$_{rest}$.}
\begin{tabular}{|l|l|}
%Row: 1
\cline{1-2}
\vbox to1.97ex{\vspace{1pt}\vfil\hbox to33.00ex{\hfil Community\hfil}\vfil} & 
\vbox to1.97ex{\vspace{1pt}\vfil\hbox to21.20ex{\hfil Features in Common\hfil}} \\

%Row: 2
\cline{1-2}
\vbox to1.70ex{\vspace{1pt}\vfil\hbox to33.00ex{\hfil \textipa{/\~i/}, \textipa{/\~a/}, \textipa{/\~u/}\hfil}\vfil} & 
\vbox to1.70ex{\vspace{1pt}\vfil\hbox to21.20ex{\hfil nasalized\hfil}\vfil} \\

%Row: 3
\cline{1-2}
\vbox to1.70ex{\vspace{1pt}\vfil\hbox to33.00ex{\hfil /{\textipa{\~i}\textlengthmark}/, %%@
/{\textipa{\~a}\textlengthmark}/, /{\textipa{\~u}\textlengthmark}/\hfil}\vfil} & 
\vbox to1.70ex{\vspace{1pt}\vfil\hbox to21.20ex{\hfil long, nasalized\hfil}\vfil} \\

%Row: 4
\cline{1-2}
\vbox to1.70ex{\vspace{1pt}\vfil\hbox to33.00ex{\hfil /{i\textlengthmark}/, /{u\textlengthmark}/, %%@
/{a\textlengthmark}/, /{o\textlengthmark}/, /{e\textlengthmark}/\hfil}\vfil} & 
\vbox to1.70ex{\vspace{1pt}\vfil\hbox to21.20ex{\hfil long\hfil}\vfil} \\

\cline{1-2}
\end{tabular}
\label{rest}
\end{table} 

\begin{table}\centering
\caption{Some of the vowel communities obtained from VoNet$_{rest^\prime}$. Comma-separated entries represent the %%@
features that are in use from the three respective classes namely the height, the backness, and the roundedness.}
\begin{tabular}{|l|l|}
%Row: 1
\cline{1-2}
\vbox to1.97ex{\vspace{1pt}\vfil\hbox to21.00ex{\hfil Community\hfil}\vfil} & 
\vbox to1.97ex{\vspace{1pt}\vfil\hbox to40.20ex{\hfil Features in Common\hfil}} \\

%Row: 2
\cline{1-2}
\vbox to1.70ex{\vspace{1pt}\vfil\hbox to21.00ex{\hfil /i/, \textipa{/\~i/}\hfil}\vfil} & 
\vbox to1.70ex{\vspace{1pt}\vfil\hbox to40.20ex{\hfil high, front, unrounded\hfil}\vfil} \\

%Row: 3
\cline{1-2}
\vbox to1.70ex{\vspace{1pt}\vfil\hbox to21.00ex{\hfil /a/, \textipa{/\~a/}\hfil}\vfil} & 
\vbox to1.70ex{\vspace{1pt}\vfil\hbox to40.20ex{\hfil low, central, unrounded\hfil}\vfil} \\

%Row: 4
\cline{1-2}
\vbox to1.70ex{\vspace{1pt}\vfil\hbox to21.00ex{\hfil /u/, \textipa{/\~u/}\hfil}\vfil} & 
\vbox to1.70ex{\vspace{1pt}\vfil\hbox to40.20ex{\hfil high, back, rounded\hfil}\vfil} \\

\cline{1-2}
\end{tabular}
\label{restd}
\end{table}

Tables~\ref{assort} ,~\ref{rest} and~\ref{restd} indicate that the communities in VoNet$_{assort}$ are formed based on %%@
the principle of perceptual contrast whereas the formation of the communities in VoNet$_{rest}$ as well as %%@
VoNet$_{rest^\prime}$ is largely governed by feature economy. We dedicate the rest of this section mainly to verify %%@
the above argument. For this reason we present a detailed study of the co-occurrence principles of the communities %%@
obtained from VoNet$_{assort}$, VoNet$_{rest}$, and VoNet$_{rest^\prime}$. In each case we compare the results with %%@
those of VoNet$_{rand}$ obtained from Algorithm~1. 

\subsection{Co-occurrence Principles of the Communities of VoNet$_{assort}$}

\begin{figure}
%\begin{center}
\centerline{\psfig{file=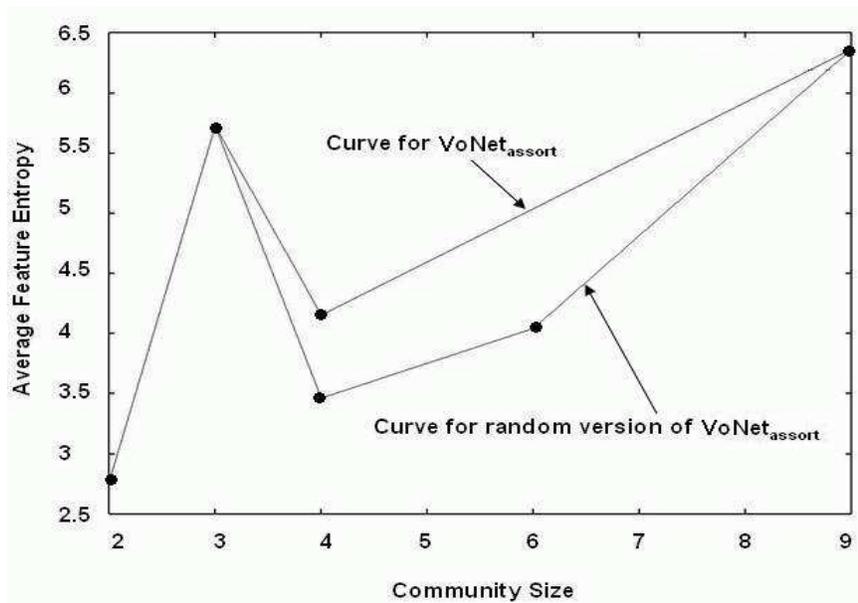,width=4.5in}}
\caption{Curves showing the average feature entropy of the communities of a particular size versus the community size %%@
for VoNet$_{assort}$ as well as its random counterpart.} 
\label{assortdst}
%\end{center}
\end{figure}

We apply the community-finding algorithm (discussed earlier) on VoNet$_{rand}$ in order to obtain the assortative %%@
communities similarly as outlined for VoNet. Figure~\ref{assortdst} illustrates, for all the communities obtained from %%@
the clustering of VoNet$_{assort}$ and its random version, the average feature entropy exhibited by the communities of %%@
a particular size\footnote{Let there be $n$ communities of a particular size $k$ picked up at various thresholds. The %%@
average feature entropy of the communities of size $k$ is therefore $\frac{1}{n}{\sum_{i=1}^n{F_{E_i}}}$ where %%@
$F_{E_i}$ signifies the feature entropy of the $i^{th}$ community.} (y-axis) versus the community size (x-axis). 

\begin{figure}
%\begin{center}
\centerline{\psfig{file=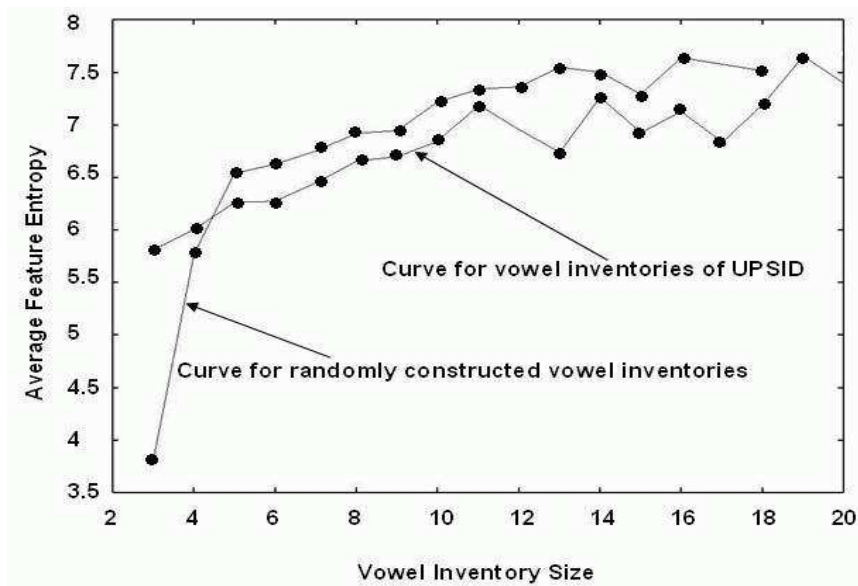,width=4.5in}}
\caption{Curves showing the average feature entropy of the vowel inventories of a particular size versus the inventory %%@
size. The two different plots compare the average feature entropy of the inventories obtained from UPSID with that of %%@
the randomly constructed ones.} 
\label{invcomp}
%\end{center}
\end{figure}

A closer inspection of Figure~\ref{assortdst} immediately reveals that the feature entropy exhibited by the %%@
communities of VoNet$_{assort}$ is higher as compared to the random version of the same. The two curves finally %%@
intersect due to the formation of a single giant component, which is similar for the real and the random edition of %%@
VoNet$_{assort}$. Nevertheless, the data points that appear on these curves are fairly less in number and hence %%@
Figure~\ref{assortdst} alone is not sufficient enough to establish that the communities in VoNet$_{assort}$ are formed %%@
based on the principle of perceptual contrast. Another possible way to investigate the problem would be to look into %%@
the co-occurrence principles of the smaller vowel inventories (of size $\le$ 4) since they mostly comprise of the %%@
members belonging to the assortative vowel communities. Table~\ref{proof} for instance, shows the number of %%@
occurrences of the members of the community formed by /$i$/, /$a$/, and /$u$/, as compared to the average occurrence %%@
of other vowels, in the inventories of size 3 and 4. The figures in the table points to the fact that the smaller %%@
inventories can be assumed to be good representatives of the assortative vowel communities. We therefore compare the %%@
average feature entropy of these inventories as a whole with their random counterparts (obtained from Algorithm~1). %%@
Figure~\ref{invcomp} illustrates the result of this comparison. The curves depict the average feature entropy of the %%@
vowel inventories of a particular size (y-axis) versus the inventory size (x-axis). The two different plots compare %%@
the average feature entropy of the inventories obtained from UPSID with that of the randomly constructed ones. The %%@
figure clearly shows that the average feature entropy of the vowel inventories of UPSID is substantially higher for %%@
inventory size 3 and 4 than that of those constructed randomly.    

% Table created by WinTeX 2000: 6 Columns x 3 Rows.
\begin{table}\centering
\caption{Frequency of occurrence of the members of the community /$i$/, /$a$/, and /$u$/, as compared to the frequency %%@
occurrence of other vowels, in smaller inventories. The last column indicates the average number of times that a vowel %%@
other than /$i$/, /$a$/, and /$u$/ occurs in the inventories of size 3 and 4.}
\begin{tabular}{|l|l|l|l|l|l|}
%Row: 1
\cline{1-6}
\vbox to1.88ex{\vspace{1pt}\vfil\hbox to8.60ex{\hfil Inv. Size\hfil}} & 
\vbox to1.88ex{\vspace{1pt}\vfil\hbox to11.20ex{\hfil No. of Invs.\hfil}} & 
\vbox to1.88ex{\vspace{1pt}\vfil\hbox to7.80ex{\hfil Occ. /i/\hfil}} & 
\vbox to1.88ex{\vspace{1pt}\hbox to8.40ex{\hfil Occ. /a/\hfil}\vfil} & 
\vbox to1.88ex{\vspace{1pt}\hbox to8.00ex{\hfil Occ. /u/\hfil}\vfil} & 
\vbox to1.88ex{\vspace{1pt}\hbox to23.40ex{\hfil Avg. Occ. other vowels\hfil}\vfil} \\

%Row: 2
\cline{1-6}
\vbox to1.88ex{\vspace{1pt}\vfil\hbox to8.60ex{\hfil 3\hfil}} & 
\vbox to1.88ex{\vspace{1pt}\vfil\hbox to11.20ex{\hfil 23\hfil}} & 
\vbox to1.88ex{\vspace{1pt}\vfil\hbox to7.80ex{\hfil 15\hfil}} & 
\vbox to1.88ex{\vspace{1pt}\hbox to8.40ex{\hfil 21\hfil}\vfil} & 
\vbox to1.88ex{\vspace{1pt}\hbox to8.00ex{\hfil 12\hfil}\vfil} & 
\vbox to1.88ex{\vspace{1pt}\hbox to23.40ex{\hfil 3\hfil}\vfil} \\

%Row: 3
\cline{1-6}
\vbox to1.88ex{\vspace{1pt}\vfil\hbox to8.60ex{\hfil 4\hfil}} & 
\vbox to1.88ex{\vspace{1pt}\vfil\hbox to11.20ex{\hfil 25\hfil}} & 
\vbox to1.88ex{\vspace{1pt}\vfil\hbox to7.80ex{\hfil 19\hfil}} & 
\vbox to1.88ex{\vspace{1pt}\hbox to8.40ex{\hfil 24\hfil}\vfil} & 
\vbox to1.88ex{\vspace{1pt}\hbox to8.00ex{\hfil 11\hfil}\vfil} & 
\vbox to1.88ex{\vspace{1pt}\hbox to23.40ex{\hfil 3\hfil}\vfil} \\

\cline{1-6}
\end{tabular}
\label{proof}
\end{table}

The results presented in Figures~\ref{assortdst}~and~\ref{invcomp} together confirms that the assortative vowel %%@
communities are formed based on the principle of maximal perceptual contrast.
 
\subsection{Co-occurrence Principles of the Communities of VoNet$_{rest}$}

\begin{figure}
%\begin{center}
\centerline{\psfig{file=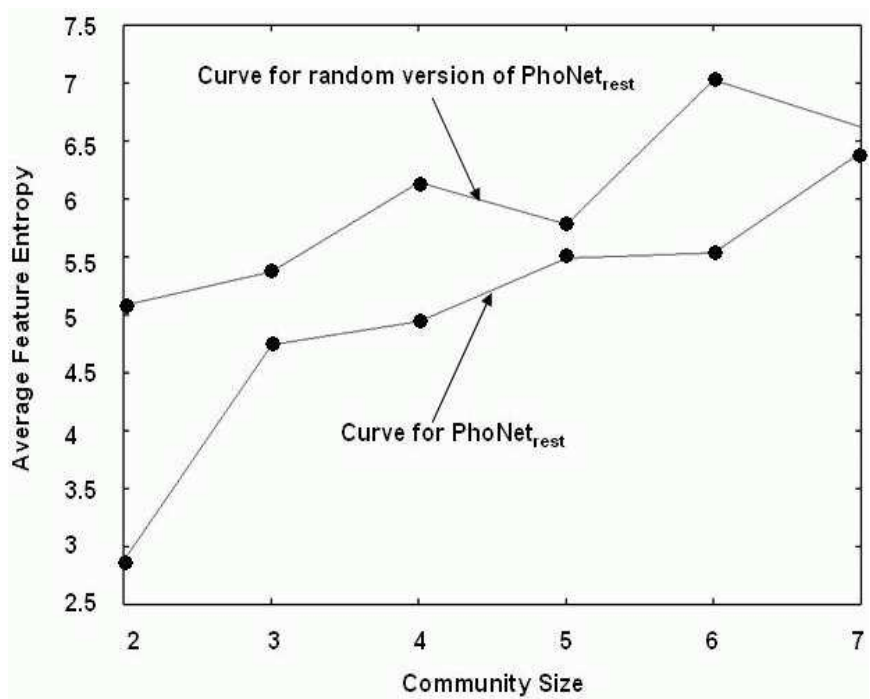,width=4.5in}}
\caption{Curves showing the average feature entropy of the communities of a particular size versus the community size %%@
for VoNet$_{rest}$ as well as its random counterpart.} 
\label{restent}
%\end{center}
\end{figure}

In this section, we investigate whether or not the communities obtained from VoNet$_{rest}$ are better in terms of %%@
feature entropy than they would have been, if the vowel inventories had evolved just by chance. We construct the %%@
random edition of VoNet$_{rest}$ from VoNet$_{rand}$ and apply the community-finding algorithm on it so as to obtain %%@
the  communities. Figure~\ref{restent} illustrates, for all the communities obtained from the clustering of %%@
VoNet$_{rest}$ and its random version, the average feature entropy exhibited by the communities of a particular size %%@
(y-axis) versus the community size (x-axis). The curves in the figure makes it quite clear that the average feature %%@
entropy exhibited by the communities of VoNet$_{rest}$ are substantially lower than that of their random counterpart %%@
(especially for a community size $\le 7$). As the community size increases, the difference in the average feature %%@
entropy of the communities of VoNet$_{rest}$ and its random version gradually diminishes. This is mainly because of %%@
the formation of a single giant community, which is similar for the real and the random versions of VoNet$_{rest}$. 

The above result indicate that the driving force behind the formation of the communities of VoNet$_{rest}$ is the %%@
principle of feature economy. It is important to mention here that the larger vowel inventories, which are usually %%@
comprised of the communities of VoNet$_{rest}$, also exhibit feature economy to a large extent. This is reflected %%@
through Figure~\ref{invcomp} where all the real inventories of size~$\ge$~5 have a substantially lower average feature %%@
entropy than that of the randomly generated ones.

\subsection{Co-occurrence Principles of the Communities of VoNet$_{rest^\prime}$}

\begin{figure}
%\begin{center}
\centerline{\psfig{file=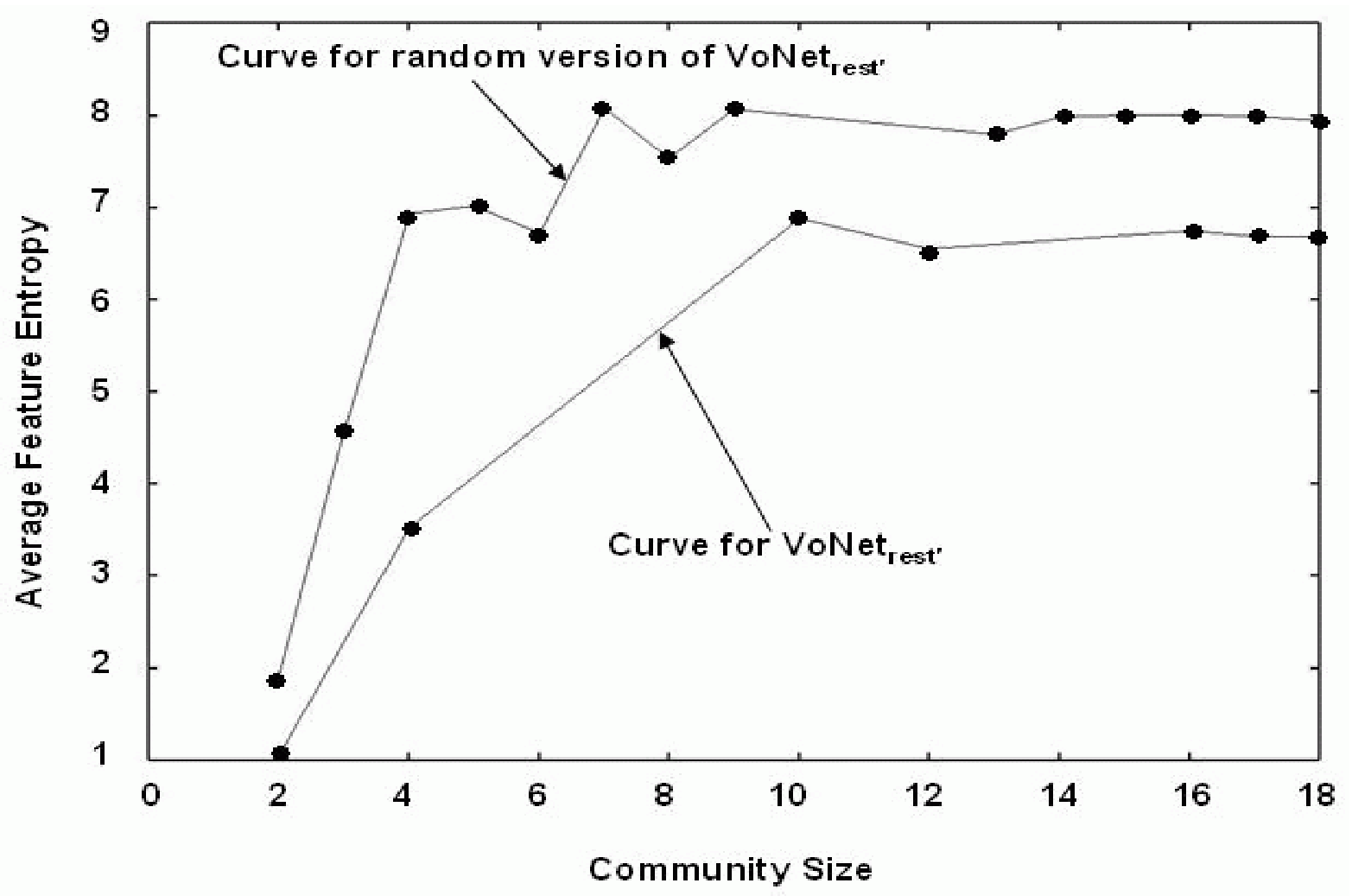,width=4.5in}}
\caption{Curves showing the average feature entropy of the communities of a particular size versus the community size %%@
for VoNet$_{rest^\prime}$ as well as its random counterpart.} 
\label{restdent}
%\end{center}
\end{figure}

In this section we compare the feature entropy of the communities obtained from VoNet$_{rest^\prime}$ with that of its %%@
random counterpart (constructed from VoNet$_{rand}$). Figure~\ref{restdent} shows the the average feature entropy %%@
exhibited by the communities of a particular size (y-axis) versus the community size (x-axis) for both the real and %%@
the random version of VoNet$_{rest^\prime}$. The curves in the figure makes it quite clear that the average feature %%@
entropy exhibited by the communities of VoNet$_{rest^\prime}$ are substantially lower than that of the random ones. %%@
This result immediately reveals that it is again feature economy that plays a key role in the emergence of the %%@
communities of VoNet$_{rest^\prime}$.

\section{Conclusion}\label{conc}

In this paper we explored the co-occurrence principles of the vowels, across the inventories of the world's languages. %%@
In order to do so we started with a concise review of the available literature on vowel inventories. We proposed an %%@
automatic procedure to capture the co-occurrence patterns of the vowels across languages. We also discussed the notion %%@
of feature entropy, which immediately allows us to validate the explanations of the organizational principles of the %%@
vowel inventories furnished by the earlier researchers. 

Some of our important findings from this work are, 

\begin{itemize}
\item The smaller vowel inventories (corresponding to the communities of\\ VoNet$_{assort}$) tend to be organized %%@
based on the principle of maximal perceptual contrast;
\item On the other hand, the larger vowel inventories (mainly comprising of the communities of VoNet$_{rest}$) reflect %%@
a considerable extent of feature economy;
\item Co-occurrences based on robustness are prevalent across vowel inventories (captured through the communities of %%@
VoNet$_{rest^\prime}$) and their emergence is again a consequence of feature economy.  
\end{itemize} 

Until now, we have mainly emphasized on analyzing the co-occurrence principles of the vowel inventories of the world's %%@
languages. An issue that draws attention is how the forces of perceptual contrast and feature economy have interacted %%@
causing the emergence of the human vowel systems. One possible way to answer this question is by having a growth model %%@
for the network, where the growth takes place owing to the optimization of a function (see~\cite{Cancho:01} for a %%@
reference), which involves the above forces and also accounts for the observed regularities displayed by the vowel %%@
inventories. It would be worthwhile to mention here that though most of the mechanisms of network growth rely on %%@
preferential attachment-based rules~\cite{Albert:99}, yet there are scenarios which suggest that additional optimizing %%@
constraints need to be imposed on the evolving network so as to match its emergent properties with empirical %%@
data~\cite{Sole:02,Vesp:03}. Such a growth model based on some optimization technique can then shed enough light on %%@
the real dynamics that went on in the evolution of the vowel inventories. We look forward to develop the same as a %%@
part of our future work.


\begin{thebibliography}{00}

\bibitem{Albert:99}
Barab{\'a}si, A.-L. and Albert, R.,
\newblock Emergence of scaling in random networks, {\em Science} {\bf 286}, 509-–512,
\newblock (1999).

\bibitem{Boer:00}
de Boer, B.,
\newblock Self-organisation in vowel systems, {\em Journal of Phonetics}, {\bf 28}(4), 441--465,
\newblock (2000).

\bibitem{Boersma:98}
Boersma, P.,  
\newblock {\em Functional phonology}, Doctoral thesis, University of Amsterdam, The Hague:
Holland Academic Graphics,
\newblock (1998).

\bibitem{Cancho:01}
Ferrer i Cancho, R. and Sol{\'e}, R. V.,
\newblock Optimization in complex networks, {\em arXiv:cond-mat/0111222},
\newblock (2001).

\bibitem{Choudhury:06}
Choudhury, M., Mukherjee, A., Basu, A. and Ganguly, N.,
\newblock Analysis and synthesis of the distribution of consonants over languages: A complex network approach, {\em %%@
Proceedings of COLING--ACL}, 128--135, Sydney, Australia,
\newblock (2006).

\bibitem{Clements:04}
Clements, N.,
\newblock  Features and sound inventories, {\em Symposium on Phonological Theory: Representations and Architecture}, %%@
CUNY,
\newblock (2004).

\bibitem{Groot:31}
de Groot, A. W., 
\newblock Phonologie und Phonetik als funktionswissenschaften, {\em Travaux du Cercle Linguistique de}, {\bf 4}, %%@
116--147,
\newblock (1931).

\bibitem{Hinskens:03}
Hinskens, F. and Weijer, J.,
\newblock Patterns of segmental modification in consonant inventories: A cross-linguistic study, {\em Linguistics}, %%@
{\bf 41}, 6.
\newblock (2003).

\bibitem{Jakobson:41}
Jakobson, R., 
\newblock {\em Kindersprache, aphasie und allgemeine lautgesetze}, (Uppsala, 1941), reprinted in {\em Selected %%@
Writings I. Mouton}, (The Hague, 1962), 328–-401.

\bibitem{Ke:03}
Ke, J., Ogura, M., and Wang, W.S.-Y., 
\newblock Optimization models of sound systems using genetic algorithms, {\em Computational Linguistics}, {\bf 29}(1), %%@
1--18,
\newblock (2003).

\bibitem{Ladefoged:96}
Ladefoged, P. and Maddieson, I.,  
\newblock {\em Sounds of the world's languages},
\newblock (Oxford: Blackwell, 1996).

\bibitem{Lindblom:72}
 Liljencrants, J. and Lindblom, B.,
\newblock Numerical simulation of vowel quality systems: the role of perceptual contrast, {\em Language}, {\bf 48}, %%@
839--862,
\newblock (1972).

\bibitem{Lindblom:86}
Lindblom, B.,
\newblock Phonetic universals in vowel systems, {\em Experimental Phonology}, 13--44,
\newblock (1986).

\bibitem{Lindblom:88}
Lindblom, B. and Maddieson, I.,
\newblock Phonetic universals in consonant systems, {\em Language, Speech, and Mind}, Routledge, London, 62--78,
\newblock (1988).

\bibitem{Maddieson:84}
Maddieson, I.,
\newblock {\em Patterns of sounds},
\newblock (Cambridge University Press, Cambridge, 1984.)

\bibitem{Martinet:55}
Martinet, A.,
\newblock {\em {\`E}conomie des changements phon{\'e}tiques},
\newblock (Berne: A. Francke, 1955).

\bibitem{Mukherjee:06}
Mukherjee, A., Choudhury, M., Basu, A. and Ganguly, N.,
\newblock Modeling the co-occurrence principles of the consonant inventories: A complex network approach, {\em %%@
arXiv:physics/0606132 (preprint)},
\newblock (2006). 

\bibitem{Mukh:06}
Mukherjee, A., Choudhury, M., Basu, A. and Ganguly, N.,
\newblock Self-organization of the Sound Inventories: Analysis and Synthesis of the Occurrence and Co-occurrence %%@
Networks of Consonants. {\em arXiv:physics/0610120 (preprint)}, 
\newblock (2006). 

\bibitem{Newman:03}
Newman, M. E. J.,
\newblock The structure and function of complex networks, {\em SIAM Review}, {\bf 45}, 167--256, 
\newblock (2003).

\bibitem{Rad:03}
Radicchi, F., Castellano, C., Cecconi, F., Loreto, V., and Parisi, D., 
\newblock Defining and identifying communities in networks, {\em PNAS}, {\bf 101}(9), 2658--2663,
\newblock (2003).

\bibitem{Schwartz:97}
Schwartz, J-L., Bo$\ddot{e}$, L-J., Vall{\'e}e, N., and Abry, C.,
\newblock The dispersion-focalization theory of vowel systems, {\em Journal of Phonetics}, {\em 25}, 255--286.
\newblock (1997).

\bibitem{Shan:49}
Shannon, C. E., and Weaver, W.,
\newblock {\em The mathematical theory of information}, 
\newblock (Urbana: University of Illinois Press, 1949).

\bibitem{Sole:02}
Sol{\'e}, R. V., Pastor-Satorras, R., Smith, E. and Kepler, T.,
\newblock A model of large-scale proteome evolution, {\em Adv. Complex Syst.}, {\bf 5}, 43--54,
\newblock (2002).

\bibitem{Vesp:03}
V{\'a}zquez, A., Flammini, A., Maritan, A., and Vespignani, A.,
\newblock Modeling of protein interaction networks, {\em Complexus}, {\bf 1}, 38--44,
\newblock (2003).

\bibitem{Wang:68}
Wang, W. S.-Y., 
\newblock The basis of speech. Project on linguistic analysis reports, (University of California, Berkeley, 1968), %%@
reprinted in {\em The Learning of Language},
\newblock (1971).

\end{thebibliography}
\end{document}